\documentclass[aps,prd,reprint,superscriptaddress,floatfix,nofootinbib,amsmath,amssymb,altaffilletter,preprintnumbers]{revtex4-2}

\usepackage[T1]{fontenc}
\usepackage{bm}
\usepackage{cancel}
\usepackage{relsize}
\usepackage[dvipsnames]{xcolor}
\usepackage{float}
\usepackage{genyoungtabtikz}
\usepackage{youngtab}

\usepackage{caption}
\usepackage{subcaption}

\usepackage[colorlinks=true, pdfstartview=FitV, breaklinks=true]{hyperref}
\hypersetup{urlcolor=BlueViolet, citecolor=Plum, linkcolor=PineGreen}
\usepackage[capitalise]{cleveref}
\crefname{appendix}{App.}{Apps.}

\newcommand{\unimelb}{\affiliation{ARC Centre of Excellence for Dark Matter Particle Physics,\\School of Physics, The University of Melbourne, Victoria 3010, Australia}}
\newcommand{\umn}{\affiliation{School of Physics and Astronomy, University of Minnesota, Minneapolis, Minnesota 55455, USA}}

\begin{document}
\preprint{UMN-TH-4301/23}
\title{\texorpdfstring{A Common Origin for the QCD Axion and Sterile Neutrinos\\ from $SU(5)$ Strong Dynamics}{A Common Origin for the QCD Axion and Sterile Neutrinos from SU(5) Strong Dynamics}}

\author{Peter Cox}
\email{peter.cox@unimelb.edu.au}
\unimelb

\author{Tony Gherghetta}
\email{tgher@umn.edu}
\umn

\author{Arpon Paul}
\email{paul1228@umn.edu}
\umn


\begin{abstract}
We identify the QCD axion and right-handed (sterile) neutrinos as bound states of an $SU(5)$ chiral gauge theory with Peccei-Quinn (PQ) symmetry arising as a global symmetry of the strong dynamics. The strong dynamics is assumed to spontaneously break the PQ symmetry, producing a high-quality axion and naturally generating Majorana masses for the right-handed neutrinos at the PQ scale. The composite sterile neutrinos can directly couple to the left-handed (active) neutrinos, realizing a standard see-saw mechanism. Alternatively, the sterile neutrinos can couple to the active neutrinos via a naturally small mass mixing with additional elementary states, leading to light sterile neutrino eigenstates. The $SU(5)$ strong dynamics therefore provides a common origin for a high-quality QCD axion and sterile neutrinos.
\end{abstract}


\maketitle

\section{Introduction}
\label{sec:intro}

The QCD axion and right-handed neutrinos are well-motivated new particle states beyond the Standard Model (SM). The QCD axion elegantly solves the strong CP problem via the Peccei-Quinn mechanism~\cite{Peccei:1977hh}, where an (anomalous) $U(1)_{\rm PQ}$ symmetry is spontaneously broken, giving rise to a pseudo-Nambu-Goldstone boson~\cite{Weinberg:1977ma,Wilczek:1977pj} that dynamically cancels the strong CP phase. Experimental bounds limit the $U(1)_{\rm PQ}$ symmetry breaking scale to be $f_{\rm PQ}\gtrsim 10^8$~GeV~\cite{Chang:2018rso}. The QCD axion can also provide the missing dark matter component of the Universe~\cite{Preskill:1982cy,Abbott:1982af,Dine:1982ah}, thereby solving two problems in the Standard Model. However, {\it explicit} violation of the $U(1)_{\rm PQ}$ global symmetry must be entirely dominated by QCD dynamics, with any other violation, in particular from gravity, highly suppressed~\cite{Georgi:1981pu,Holman:1992us,Kamionkowski:1992mf,Barr:1992qq,Ghigna:1992iv}. An attractive solution to this axion quality problem is to realize the PQ symmetry as an accidental symmetry in the low-energy theory, similar to baryon and lepton number in the Standard Model. In particular, if new strong dynamics is introduced around the PQ breaking scale, then gauge and Lorentz symmetry can be used to accidentally preserve $U(1)_{\rm PQ}$ up to very high dimension terms in the Lagrangian~\cite{Randall:1992ut}.

The seesaw mechanism~\cite{Minkowski:1977sc,Yanagida:1979as,GellMann:1980vs,Glashow:1979nm,Mohapatra:1979ia} provides a similarly elegant explanation of the hierarchy between the masses of neutrinos and charged leptons. Assuming order one Yukawa couplings, this is simply achieved by introducing right-handed neutrinos with masses $\gtrsim 10^{10}$~GeV. As is well-known, this mass scale is similar to the PQ scale and the two can be related~\cite{Mohapatra:1982tc,Langacker:1986rj}. Given this coincidence of scales, and the possibility of realizing the PQ symmetry as an accidental symmetry, we seek a solution that relates these two scales via strong dynamics. The strong dynamics also has the advantage of naturally generating the PQ scale via dimensional transmutation, obviating the need to introduce explicit mass scales in the scalar potentials that are normally used to UV complete the axion Lagrangian, as in the KSVZ or DFSZ scenarios.

A particularly interesting strong dynamics based on an $SU(5)$ gauge theory with massless, chiral fermions was recently considered in Ref.~\cite{Gavela:2018paw} to realize a high-quality QCD axion. The PQ symmetry was identified as a global symmetry of the strong dynamics and assumed to be spontaneously broken at a scale $f_{\rm PQ}$. This realises a low-energy QCD axion as a composite pseudo Nambu-Goldstone (NG) boson, thereby solving the strong CP problem in a similar manner to the original dynamical axion~\cite{Kim:1984pt,Choi:1985cb}. Furthermore, the local gauge and Lorentz symmetry accidentally preserves the $U(1)_{\rm PQ}$ global symmetry up to dimension nine terms in the low-energy effective Lagrangian, thereby ameliorating the axion quality problem.

In this paper, we build upon the $SU(5)$ model in Ref.~\cite{Gavela:2018paw} to realize both a high-quality QCD axion and right-handed (sterile) neutrinos as bound states of the {\it same} UV dynamics. In particular, we show that the spontaneous breakdown of the PQ symmetry leads to QCD singlet states with Majorana masses of order the PQ scale, $f_{\rm PQ}$, which can be identified as composite sterile neutrinos. To generate the left-handed (active) neutrino masses, the composite sterile neutrinos are then coupled either directly or indirectly to the active neutrinos, realizing heavy or light sterile neutrino mass eigenstates, respectively. 

In the case where the composite sterile neutrinos couple directly to the active neutrinos (with a dimension-seven Higgs--fermion coupling), light Majorana active neutrinos are obtained via a see-saw mechanism. The interaction is generated by integrating out PQ-charged scalar fields in a UV completion. Importantly, the quality of the PQ symmetry is not affected. Alternatively, an indirect coupling to the active neutrinos can occur via a naturally generated small mass mixing between an elementary, right-handed neutrino and the composite sterile neutrinos. 
This leads to sterile states with naturally suppressed (sub-TeV) Majorana masses and, depending on the scale of the UV completion, can realize the neutrinos as pseudo-Dirac states~\cite{Arkani-Hamed:1998wff,Gherghetta:2003hf}. The strong dynamics, therefore, plays a pivotal role not only in addressing the axion quality problem, but also in relating the axion and neutrino masses.

The composite sterile neutrinos share features similar to those previously studied in Refs.~\cite{Arkani-Hamed:1998wff, Gherghetta:2003hf, Agashe:2015izu, Chacko:2020zze, Chakraborty:2021fkp}, however the connection with a composite QCD axion was not previously considered. The chiral UV gauge theory also provides an explicit 4D realization of the holographic 5D setups considered in Refs.~\cite{Cox:2019rro,Bonnefoy:2020llz, Cox:2021lii}, which solved the axion quality problem with a composite axion and partial compositeness in the SM fermion sector. Finally, previous work in Refs.~\cite{Redi:2016esr,Lillard:2018fdt,Vecchi:2021shj, Contino:2021ayn} also addressed the axion quality problem with an accidental PQ symmetry of strong dynamics, although without any connection to neutrino masses.

The outline of our paper is as follows. In \cref{sec:SU5} we review the matter content of the $SU(5)$ gauge theory, together with the global symmetry structure and IR dynamics. We then discuss the resulting bound state spectrum, which includes a composite, high-quality axion as well as QCD singlet bound states that are identified as composite right-handed neutrinos. The generation of neutrino masses is discussed in \cref{sec:NeutrinoMass}, where we present models with both heavy and light sterile neutrino mass eigenstates. A holographic connection to the light sterile neutrino case is also discussed. Our concluding remarks are presented in \cref{sec:conc}. The Appendices contain supplementary material related to the QCD anomaly factor (\cref{app:QCDanom}), representations of the NG bosons (\cref{app:NGB-rep}), solution of the axion quality problem (\cref{app:axion-quality}), implications for axion dark matter (\cref{app:dark-matter}), and mass-mixing in the light sterile neutrino scenario (\cref{app:simplemassmixing}).

\section{\texorpdfstring{$SU(5)$}{SU(5)} Chiral Gauge Theory}
\label{sec:SU5}

\subsection{Gauge and Global Symmetries}

We consider the $SU(5)$ gauge theory of Ref.~\cite{Gavela:2018paw} which extends the SM gauge group to $SU(5)\times SU(3)_c\times SU(2)_L\times U(1)_Y$ and introduces massless, chiral (left-handed) fermions\footnote{We use two component notation throughout the paper.} transforming in the anti-fundamental ($\bf\bar 5$) and antisymmetric ($\bf 10$) representations of $SU(5)$. Under the SM gauge group, these new fermions transform in a (pseudo) real representation $\mathbf{R}_\psi$ of $SU(3)_c$ and are singlets under $SU(2)_L\times U(1)_Y$ (see \cref{tab:gauge}). All gauge anomalies vanish by construction. Later, we consider two specific realisations with $SU(3)_c$ representations $\mathbf{R}_\psi =\textbf{3}\oplus \bar{\textbf{3}}$ and $\mathbf{R}_\psi =\textbf{8}$.

\begin{table}[t]
    \caption{Representations of the chiral fermions charged under the $SU(5)\times SU(3)_c$ gauge symmetry. \label{tab:gauge}}
    \renewcommand{\arraystretch}{1.5}
    \begin{ruledtabular}
    \begin{tabular}{lcc} 
    & $SU(5)$ & $SU(3)_c$  \\ \hline
      $\psi_{\bar{5}}$ & \Yvcentermath1$\overline{
    \tiny\yng(1)}$ & $\textbf{R}_\psi$ \\
    $\psi_{10}$ & \Yvcentermath1${\tiny\yng(1,1)}$  & $\textbf{R}_\psi$ 
    \end{tabular}
    \end{ruledtabular}
\end{table}

In the limit that the QCD coupling $\alpha_s \rightarrow 0$, the $SU(5)$ gauge theory has an $SU(n_f)_{\bar{5}} \times SU(n_f)_{10}$ global symmetry, where $n_f={\rm dim\,{\bf R}}_\psi$. In addition, there is a single $SU(5)$-anomaly-free global $U(1)$ (analogous to $B-L$ symmetry in $SU(5)$ grand unified theories) for which the charges of $\psi_{\bar 5}$ and $\psi_{10}$ satisfy $Q_{\bar{5}} = -3 Q_{10}$. This is identified as the PQ symmetry. The representations of the fermions under the full flavor symmetry are shown in \cref{tab:global}. The QCD gauge group, $SU(3)_c$, is a subgroup of the non-abelian flavor symmetry, and the latter is explicitly broken for $\alpha_s \neq 0$. Importantly, the PQ symmetry is anomalous with respect to QCD (see \cref{app:QCDanom}), which will eventually lead to the composite axion obtaining a mass from non-perturbative QCD effects in the usual way. On the other hand, the fact that $U(1)_\text{PQ}$ has no $SU(5)$ anomaly is important to ensure that the axion remains light and provides a solution to the strong CP problem.

\begin{table}[t]
    \caption{Representations of the $SU(5)$ chiral fermions under the global flavor symmetry $SU(n_f)_{\bar{5}}\times SU(n_f)_{10}\times U(1)_{\rm PQ}$ (in the limit $\alpha_s\rightarrow 0$). \label{tab:global}}
    \renewcommand{\arraystretch}{1.5}
    \begin{ruledtabular}
    \begin{tabular}{lccccc} 
    & $SU(n_f)_{\bar{5}}$ & $SU(n_f)_{10}$  & $U(1)_{\rm PQ}$\\
    \colrule
    $\psi_{\bar{5}}$& \Yvcentermath1${\tiny\yng(1)}$ & \textbf{1} & -3/5 \\
     $\psi_{10}$ & \textbf{1} & $\Yvcentermath1{\tiny\yng(1)}$ &  1/5 
    \end{tabular}
    \end{ruledtabular}
\end{table}

\subsection{IR Dynamics and Symmetry Breaking}
\label{sec:symmetry-breaking}

Given the fermion content in \cref{tab:gauge}, the $SU(5)$ gauge theory is asymptotically free in the UV and becomes strongly coupled in the IR. The dynamics of strongly coupled, non-supersymmetric chiral gauge theories are not well understood. Techniques such as 't Hooft anomaly matching, large-$N$, and the $a$-theorem can be used to place restrictions on the dynamics but do not, in general, single out a unique IR phase. (For a recent discussion of the IR dynamics of $SU(N)$ theories with a single flavor of antisymmetric + anti-fundamental chiral fermions see \cite{Karasik:2022gve}.)

It was pointed out in Ref.~\cite{Gavela:2018paw} that for the current $SU(5)$ model with $n_f$ flavors it is impossible to match the $[SU(n_f)_{\bar{5}}]^3$ and $[SU(n_f)_{10}]^3$ anomalies if the $SU(5)$ confines in the IR. The global symmetry is therefore spontaneously broken by the $SU(5)$ gauge dynamics; however, there remain several possible IR phases with different unbroken global symmetry groups. Furthermore, there is the possibility of forming bilinear condensates (i.e. $\langle \psi_{\bar 5}\psi_{10} \rangle$) that dynamically break the $SU(5)$ gauge theory (see e.g.~\cite{Appelquist:2000qg}). Following \cite{Gavela:2018paw}, we assume that (i) the gauge theory confines and no bilinear condensates form, and (ii) the flavor breaking condensate preserves {\it at least} an $SU(3)$ subgroup, which is the weakly gauged $SU(3)_c$. 

While there are no fermion bilinears that are both $SU(5)$ singlets and Lorentz scalars, the flavor symmetry breaking can be achieved through the 4-fermion condensate\footnote{In fact, there are three more possible condensates to achieve the flavor symmetry breaking: $\langle \psi_{\bar 5}\psi_{\bar 5}\psi_{\bar 5}^\dagger\psi_{\bar 5}^\dagger \rangle $,  $\langle \psi_{10}\psi_{10} \psi_{10}^\dagger \psi_{10}^\dagger \rangle$, and $\langle \psi_{\bar 5}\psi_{\bar 5}^\dagger \psi_{10} \psi_{10}^\dagger \rangle $.}
\begin{equation}
    \langle \Phi_{\text{flavor}}\rangle\equiv\langle \psi_{\bar 5}\psi_{10} \psi_{10} \psi_{10} \rangle \,,
\end{equation}
which is $SU(5)$-invariant and contains an $SU(3)_c$ singlet. The breaking pattern is then
\begin{align} \label{eq:global-symmetry-breaking}
    SU(n_f)_{\bar{5}} \times & SU(n_f)_{10} \xrightarrow{\langle \Phi_{\text{flavor}}\rangle \neq 0}  \mathcal{G} \supseteq SU(3)_c \,.
\end{align}
This spontaneous breaking of the non-abelian flavor symmetry gives rise to a large number of (pseudo-)NG bosons (see \cref{app:NGB-rep}). For fermions in the $\mathbf{R}_\psi =\textbf{8}$ representation of QCD, the gauging of $SU(3)_c$ leaves no residual continuous global symmetries (besides $U(1)_{\rm PQ}$). All of these pseudo-NG bosons are therefore colored and acquire masses of order $\sim \sqrt{\alpha_s} \Lambda_5$, where $\Lambda_5$ is the typical mass scale of the composite bound states. On the other hand, for fermions in the $\mathbf{R}_\psi =\textbf{3}\oplus \bar{\textbf{3}}$ representation, the $SU(6)_{\bar{5}} \times SU(6)_{10}$ flavor symmetry is explicitly broken to $SU(3)_c \times U(1) \times U(1)'$. If these $U(1)$ symmetries are spontaneously broken (i.e. not contained in $\mathcal{G}$) there are two massless, QCD-singlet NG bosons. This latter case would therefore be excluded by cosmological bounds on the number of relativistic degrees of freedom ($N_\text{eff}$) if the $SU(5)$ sector was in equilibrium with the SM in the early universe.

The condensate $\langle \Phi_{\text{flavor}}\rangle$ has PQ charge zero and preserves $U(1)_{\rm PQ}$. The lowest dimension $SU(5)$-singlet operators that carry PQ charge contain six fermions and are dimension-9:
\begin{align} \label{eq:PQcondensate}
    \Phi_{\rm PQ,1}  &\equiv \psi_{\bar{5}}\,\psi_{\bar{5}}\,\psi_{10}\,\psi_{\bar{5}}\,\psi_{\bar{5}}\,\psi_{10} \,, \notag \\ 
    \Phi_{\rm PQ,2} &\equiv \psi_{\bar{5}}\,\psi_{10}^\dagger\,\psi_{10}^\dagger\,\psi_{\bar{5}}\,\psi_{10}^\dagger\,\psi_{10}^\dagger \,, \notag \notag \\
    \Phi_{\rm PQ,3} &\equiv \psi_{\bar{5}}\,\psi_{\bar{5}}\,\psi_{10}\,\psi_{\bar{5}}\,\psi_{10}^\dagger\,\psi_{10}^\dagger \,, \notag \\
    \Phi_{\rm PQ,4} &\equiv \psi_{\bar{5}}\,\psi_{\bar{5}}\,\psi_{\bar{5}}\,\psi_{\bar{5}}\,\psi_{\bar{5}}^\dagger\,\psi_{10}^\dagger \,.
\end{align}
These operators all have PQ charge $-2$. It is assumed that at least one of these condenses and spontaneously breaks $U(1)_\text{PQ}$, while preserving $SU(3)_c$.

\subsection{Composite axion}

The spontaneous PQ breaking by the condensate \eqref{eq:PQcondensate} gives rise to a composite pseudo-NG boson, which is the axion. We parameterise the Goldstone field containing the axion, $a$, as 
\begin{equation} \label{eq:U}
    U = e^{i \frac{a}{f_{\text{PQ}}}} \,,
\end{equation}
where $a/f_{\rm PQ} \in [0,2\pi)$. Under the PQ transformation $\psi \to e^{iQ_\psi\alpha}\psi$, with $\alpha$ an arbitrary phase parameter, the axion transforms as $a \to a + \alpha f_{PQ}$, where $f_\text{PQ}$ is the axion decay constant. This constant obeys the relation $f_\text{PQ}=\Lambda_5/g_*$, where  $g_*$ is a typical coupling between the composite bound states (with mass scale $\Lambda_5$), satisfying $1\lesssim g_* \lesssim 4\pi$.

As usual, QCD instantons generate a potential for the axion, providing a dynamical solution to the strong CP problem. The axion mass is then given by the standard expression~\cite{GrillidiCortona:2015jxo,Gorghetto:2018ocs},
\begin{align}
    m_a &\approx 5.7\times 10^{-6}\,\text{eV}\left( \frac{ 10^{12}\,\text{GeV}}{f_a} \right) \,, \label{eq:ma}
\end{align}
where $f_a \equiv f_{\rm PQ} / \mathcal{N}$, with $\mathcal{N}$ the anomaly coefficient (see \cref{app:QCDanom}).

The solution to the strong CP problem is spoiled if there are additional sources of explicit PQ violation, which will modify the axion potential. As is well known~\cite{Holman:1992us,Kamionkowski:1992mf,Barr:1992qq,Ghigna:1992iv}, gravity is not expected to preserve the global PQ symmetry and may induce higher-dimensional, Planck scale suppressed operators that contain the PQ-charged fermions. Below the scale where $SU(5)$ confines, these operators give additional contributions to the axion potential. Importantly, the combination of Lorentz symmetry and the $SU(5)$ gauge symmetry restricts these operators to have dimension-9 or greater, such that $U(1)_{\rm PQ}$ remains an approximate accidental symmetry at low energies. Planck scale induced contributions to the axion potential are then sufficiently suppressed provided that $f_{a}\lesssim10^9$\,GeV~\cite{Gavela:2018paw} (see \cref{app:axion-quality}). This bound also has consequences for axion dark matter, which is further discussed in \cref{app:dark-matter}.

\subsection{Composite Fermion Bound States}
\label{sec:bound-states}

In addition to the composite axion, there are massive fermionic bound states. As we now show, these include SM singlets that, as will be discussed further in \cref{sec:NeutrinoMass}, can act as massive sterile neutrinos. We restrict our discussion to the 3-fermion, spin-$\frac{1}{2}$ bound states, for which there are two distinct $SU(5)$ singlets,
\begin{align}
    \Psi_1 &\equiv \psi^\dagger_{\bar{5}} \psi^\dagger_{\bar{5}} \psi^\dagger_{10}\,, \label{eq:Psi1} \\
    \Psi_2 &\equiv \psi^\dagger_{\bar{5}} \psi_{10}  \psi_{10} \,, \label{eq:Psi2}
\end{align}
where we have written the bound states as right-handed Weyl fermions for convenience. These bound states all have PQ charge $+1$, and due to the spontaneous breaking of $U(1)_\text{PQ}$ by the condensate \eqref{eq:PQcondensate} obtain Majorana masses of order the resonance scale $\Lambda_5$. 

The bound states decompose into irreducible representations of $SU(3)_c$ that depend on the representation $\textbf{R}_\psi$ of the constituent fermions. For example, with $\mathbf{R}_\psi = \mathbf{3}\oplus\bar{\mathbf{3}}$, we have:
\begin{align}
    \mathbf{3}\otimes\mathbf{3}\otimes\mathbf{3} &= \mathbf{1} \oplus  \dots \, \nonumber\\
    \bar{\mathbf{3}}\otimes\bar{\mathbf{3}}\otimes\bar{\mathbf{3}} &= \mathbf{1} \oplus  \dots \, \\
    \mathbf{3}\otimes\mathbf{3}\otimes\bar{\mathbf{3}} &= 2(\mathbf{3}) \oplus \dots~~~ \text{etc.} \nonumber
\end{align}
Similarly, for $\mathbf{R}_\psi = \mathbf{8}$,
\begin{align}
    \mathbf{8}\otimes\mathbf{8}\otimes\mathbf{8} &= 2(\mathbf{1}) \oplus \dots\,
\end{align}
Note that in both cases $\Psi_1$ and $\Psi_2$ each contain two QCD singlet bound states\footnote{Note that for $\mathbf{R}_\psi = \mathbf{3}\oplus\bar{\mathbf{3}}$, the $SU(3)_c$ singlet bound states in $\Psi_2$ must have angular momentum $L=1,3,...$ for the wavefunction to be antisymmetric under fermion exchange.}, which can be identified as right-handed neutrino candidates. We denote these bound states by $N_{1,j}$ and $N_{2,j}$, where the index $j=1,2$ represents the two different $SU(3)_c$ singlets:
\begin{align}
    \Psi_1 = N_{1,1} \oplus N_{1,2} \oplus (\text{QCD colored}) \,, \label{eq:N1} \\
    \Psi_2 = N_{2,1} \oplus N_{2,2} \oplus (\text{QCD colored}) \,. \label{eq:N2}
\end{align}

It was shown in \cite {Gavela:2018paw} that the 't Hooft anomaly matching condition for $U(1)_\text{PQ}$ can be satisfied if the bound state in $\Psi_1$ that transforms in the $\mathbf{R}_\psi$ representation is massless. Thus, anomaly matching provides no guidance as to whether or not $U(1)_\text{PQ}$ is spontaneously broken when the theory confines. As discussed in \cref{sec:symmetry-breaking}, we assume that it is spontaneously broken, which provides the necessary mechanism to realize a QCD axion in the low-energy theory.

\section{Neutrino mass}
\label{sec:NeutrinoMass}

A particularly interesting feature of the model is that the same dynamics that spontaneously breaks the PQ symmetry and generates a composite, high-quality axion also produces composite, QCD singlet fermions. These can be identified as composite sterile neutrinos if there exist additional couplings that connect the strongly-interacting $SU(5)$ sector with the electroweak sector of the SM. The active neutrinos then mix with the spectrum of composite sterile neutrino states which, together with the effect of the PQ breaking condensate, leads to the generation of Majorana masses for the active neutrinos. The PQ symmetry therefore serves as a generalised lepton number. In the following sections we present two explicit realisations of this idea. First, we consider a model that, after integrating out the $SU(5)$ sector (containing the heavy sterile neutrinos), reduces to the Weinberg operator at low energies. Second, we consider an alternative model that contains elementary sterile neutrinos which mix with the composite states, resulting in light sterile mass eigenstates.

\subsection{Heavy Sterile Neutrino Model}

Before presenting a renormalisable UV model, we first consider how the $SU(5)$ and SM lepton sectors can be connected within an effective field theory (EFT) framework. The lowest dimension operators that can achieve this are dimension-7:
\begin{multline} \label{eq:LagEFT1}
    \mathcal{L}_\text{EFT} = \frac{\tilde{\xi}_{ij}}{\Lambda_L^3}L_iH \big( \psi_{\bar{5}} \psi_{\bar{5}} \psi_{10} \big)_j \\
    +  \frac{\tilde{\xi}'_{ij}}{\Lambda_L^3}L_iH \big( \psi_{\bar{5}} \psi_{10}^\dagger \psi_{10}^\dagger \big)_j +\text{h.c.} \,,
\end{multline} 
where $L_i = (\nu_{L,i}, e_{L,i})^T$ are the SM $SU(2)$ lepton doublets, which carry PQ charge $+1$, and $H$ is the Higgs doublet\footnote{Note that we suppress $SU(2)_L$ indices, i.e. $L H = \epsilon_{ab} L^a H^b$.} which obtains the VEV $\langle H\rangle \equiv \frac{1}{\sqrt{2}}( 0, v)^T$, with $v \approx 246\,\text{GeV}$. The index $j$ enumerates the $SU(3)_c$ singlets in $(\psi \psi \psi)$, with $j\in\{1,2\}$ for either $\mathbf{R}_\psi = \mathbf{3}\oplus\bar{\mathbf{3}}$ or $\mathbf{R}_\psi = \mathbf{8}$, as discussed in section \ref{sec:bound-states}. We assume that the scale $\Lambda_L$ satisfies $\Lambda_5 < \Lambda_L < M_{\rm Pl}$, with the dimensionless couplings $\tilde{\xi}_{ij},\,\tilde{\xi}'_{ij}$ allowing for flavor-dependent masses for the different neutrino flavors denoted by the index $i$.

\begin{figure*}[t]
    \centering
    \includegraphics[width = .49\linewidth]{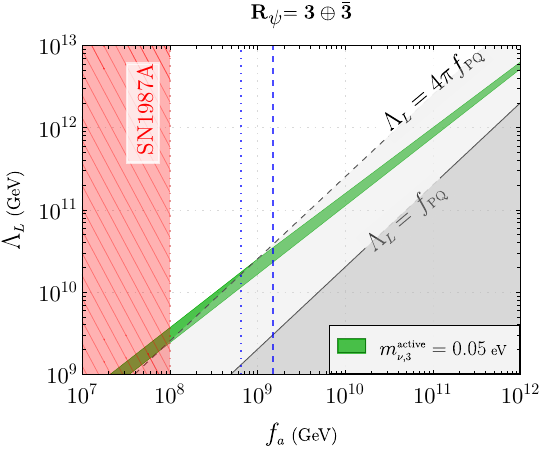} \hfill 
    \includegraphics[width = .49\linewidth]{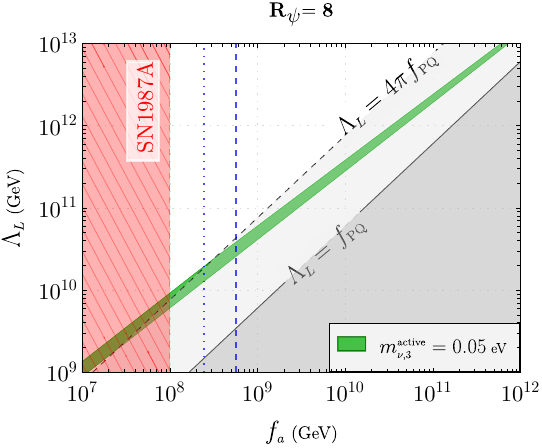} 
    \caption{EFT scale $\Lambda_L$ versus $f_a$ in the heavy sterile neutrino model for $\mathbf{R}_\psi = \mathbf{3} \oplus \mathbf{\bar{3}}$ (left) and $\mathbf{R}_\psi =\mathbf{8}$ (right). Within the green band an active neutrino mass $m_{\nu,3}^{\rm active} = 0.05\,{\rm eV}$ is obtained with $1\lesssim g_* \lesssim 4\pi$ and $\Lambda_L>\Lambda_5$, assuming $\lambda_{\nu,3} = 1$. The red shaded region is excluded by the SN1987A bound on $f_a$~\cite{Chang:2018rso}. The estimated upper limit on $f_a$ for a high-quality axion consistent with the neutron EDM bound (assuming $\left| {\text{Im}}\left(c_{\cancel{\rm PQ}}\right)\right| \gtrsim 10^{-3}$, see \cref{app:axion-quality}) is shown by the blue dashed (dotted) line for $g_* = 1 \,(4\pi)$.  The breakdown of the EFT validity when $\Lambda_L \lesssim \Lambda_5 (=g_* f_{\rm PQ})$ is depicted by the dark (light) grey shaded region for $g_* = 1 \,(4\pi)$.}
    \label{fig:LambdaL}
\end{figure*}

The relevant low-energy effective theory below the $SU(5)$ resonance scale, $\Lambda_5$, contains only the SM degrees of freedom and the pseudo-NG axion, with the PQ symmetry non-linearly realised. (The heavy composite resonances, including the singlet fermion bound states, have been integrated out.) The leading term, consistent with the symmetries, that is induced by the operators in \cref{eq:LagEFT1} is 
\begin{equation} \label{eq:LHLH_axion}
    \mathcal{L} = \xi_{ik}\xi_{jk} \left(\frac{\Lambda_5 f_{\rm PQ}^2}{\Lambda_L^3}\right)^2 \frac{1}{\Lambda_5} (L_i H) (L_j H) e^{-2ia/f_\text{PQ}} + \text{h.c.} \,,
\end{equation}
where $\xi_{ij}\simeq\mathcal{O}(1) \times \tilde{\xi}_{ij}$ and, for simplicity, we have taken $\tilde{\xi}'_{ij}=0$. The factors of $f_{\rm PQ}$ and $\Lambda_5$ have been determined using dimensional analysis, assuming the strong dynamics can be described by a single mass scale and coupling (see e.g.~\cite{Panico:2015jxa}). After electroweak symmetry breaking, the above term generates Majorana masses for the neutrinos,
\begin{equation} \label{eq:mnu1}
   m^\text{active}_{\nu,i} = \lambda_{\nu,i}\left(\frac{\Lambda_5 f_{\rm PQ}^2}{\Lambda_L^3}\right)^2 \frac{v^2}{2\Lambda_5} \,,
\end{equation}
where $\lambda_{\nu,i}$ are the eigenvalues of $\xi_{ik} \xi_{jk}$ in \eqref{eq:LHLH_axion}. Notice that the neutrino masses feature the usual see-saw factor $v^2/\Lambda_5$ (with $\Lambda_5$ identified as the scale of the heavy sterile neutrinos), but also an additional suppression by the ratio of the resonance scale to the EFT scale $\Lambda_L$. Consequently, reproducing the measured neutrino masses requires $\Lambda_5$ to be lower than the usual Type-I see-saw scale. A lower $\Lambda_5$ is also desirable to address the axion quality problem. Notice also that \cref{eq:LHLH_axion} leads to axion--neutrino couplings.

Comparing \cref{eq:mnu1,eq:ma}, the ratios between the active neutrino masses and the axion mass are estimated to be
\begin{equation}
    \frac{m^\text{active}_{\nu,i}}{m_a} \sim \frac{\lambda_{\nu,i}}{\mathcal{N}g_*^5}\left(\frac{13.2\,\Lambda_5}{\Lambda_L}\right)^6 \,.
\end{equation}
This shows that if the EFT scale is close to the PQ scale, specifically when $\Lambda_L \simeq
(13/g_*^{5/6})\,\Lambda_5$, the axion and neutrino masses are in approximately the same range.

The expression \eqref{eq:mnu1} is fitted to the observed neutrino mass spectrum to determine the viable parameter space. Assuming the neutrinos are normal ordered, and fixing $\lambda_{\nu,3}=1$, we use the mass of the heaviest neutrino to constrain $\Lambda_L$ in terms of $\Lambda_5$. The lighter neutrino masses are then simply obtained by choosing appropriate $\lambda_{\nu,1}$, $\lambda_{\nu,2}$. The combination of neutrino oscillation measurements~\cite{Esteban:2020cvm,ParticleDataGroup:2022pth} and the upper bound on the sum of neutrino masses from cosmology~\cite{Planck:2018vyg,eBOSS:2020yzd,DES:2021wwk} ($\sum m_{\nu,i} < 0.13\,\text{eV}$) leads to the $2\sigma$ range $0.05\,\text{eV} \leq m^\text{active}_{\nu,3}\leq 0.06\,\text{eV}$. This restricts the allowed values of $f_a$ and $\Lambda_L$, as shown in \cref{fig:LambdaL} for $\mathbf{R}_\psi = \mathbf{3} \oplus \mathbf{\bar{3}}$ (left panel) and $\mathbf{R}_\psi =\mathbf{8}$ (right panel). Within the green band an active neutrino mass $m_{\nu,3}=0.05\,$eV can be obtained with a strong sector coupling in the range $1 \leq g_* \leq 4\pi$. The lower edge of the band corresponds to $g_*=1$ and the upper edge to $g_*=\min(4\pi,\Lambda_L/f_{PQ})$, such that $\Lambda_L > \Lambda_5$ is always satisfied within the band. The range of $f_a$ excluded by SN1987A~\cite{Chang:2018rso} is shown in red. Values of $f_a$ to the right of the dashed ($g_*=1$) or dotted ($g_*=4\pi$) blue line are disfavoured, since Planck-suppressed contributions to the axion potential can destabilise the solution to the strong CP problem (see \cref{app:axion-quality}). 

The EFT description in \cref{eq:LagEFT1} should remain valid up to the energy scale of the composite resonances, otherwise the new degrees of freedom in the UV completion will, in general, modify the flavor and PQ symmetry breaking dynamics discussed in \cref{sec:symmetry-breaking}. (While such a scenario could also be viable, we do not consider it here.) This corresponds to the requirement $\Lambda_L > \Lambda_5$. This condition is violated in the dark (light) grey regions in \cref{fig:LambdaL} for $g_*=1\,(4\pi)$. Taking into account the lower bound on $f_a$ from SN1987A, we then find that in most of the parameter space small values of the strong sector coupling, $g_*\simeq 1$, are needed to generate the active neutrino masses in this scenario.

\subsubsection{UV completion}

\begin{figure}[t]
    \begin{subfigure}[t]{0.48\linewidth}
    \includegraphics[width = \linewidth]{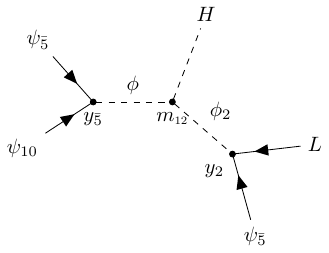}
    \caption{$\psi_{\bar 5} \psi_{\bar 5} \psi_{10} L H$}
    \end{subfigure}
    \begin{subfigure}[t]{0.48\linewidth}
    \includegraphics[width = \linewidth]{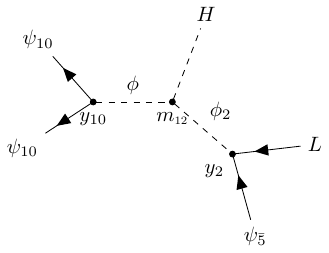}
    \caption{$\psi_{\bar 5} \psi_{10}^\dagger \psi_{10}^\dagger L H$}
    \end{subfigure}
    \caption{UV-completion of the dimension-7 operators in \cref{eq:integratedL} arising from tree-level exchange of the heavy scalars $\phi$ and $\phi_2$. (The arrows represent fermion number.)}
    \label{fig:feynModelA}
\end{figure}

A UV completion of the operators in \cref{eq:LagEFT1} can be obtained by introducing two massive complex scalar fields $\phi,\,\phi_2$. We take these fields to have masses $m_{\phi},m_{\phi_2}>\Lambda_5$, so that they do not affect the confinement and symmetry breaking of the $SU(5)$ strong dynamics discussed in \cref{sec:SU5}. In addition, the scalars do not obtain VEVs and therefore do not reintroduce the axion quality problem (or affect electroweak symmetry breaking). The relevant interaction Lagrangian is\footnote{The trilinear Higgs-scalar coupling and quartic couplings such as $|\phi|^2 |H|^2$, $|\phi_2|^2 |H|^2$ induce large radiative corrections to the Higgs mass which clearly need to be tuned. We do not address the hierarchy problem in this paper. The quartic couplings also ensure the scalar potential is bounded from below.}
\begin{multline}
    \mathcal{L} = y_{\bar 5} \psi_{\bar{5}}\psi_{10}\phi + \frac{1}{2} y_{10} \psi_{10} \psi_{10} \phi^\dagger + y_2 L \psi_{\bar{5}}\phi_2^\dagger \\
    + m_{12} \phi \phi_2^\dagger H^\dagger + {\rm h.c.} \,,
    \label{eq:UVheavy}
\end{multline}
where $y_{\bar 5},y_{10},y_2$ are dimensionless couplings and $m_{12}~\lesssim~m_\phi,~ m_{\phi_2}$ is a mass parameter. Note that we have suppressed the indices on the Yukawa couplings $y_{\bar{5}}$, $y_{10}$ that enumerate the different $SU(3)_c$ contractions, as well as the generation indices of the lepton doublet and coupling $y_2$. The charges of the fields are listed in \cref{tab:UVmodel-heavy}. Integrating out $\phi$ and $\phi_2$, as shown diagrammatically in \cref{fig:feynModelA}, yields the effective Lagrangian
\begin{multline}
    \mathcal{L}_\text{eff} = \frac{y_{\bar 5} y_2 m_{12}}{m_{\phi}^2 m_{\phi_2}^2} H (L \psi_{\bar{5}} ) (\psi_{\bar{5}} \psi_{10}) \\
    + \frac{1}{2}\frac{y_{10} y_2 m_{12}}{m_{\phi}^2 m_{\phi_2}^2} H (L \psi_{\bar{5}} ) (\psi_{10}^\dagger \psi_{10}^\dagger) + {\rm h.c.} \,. \label{eq:integratedL}
\end{multline}
The two terms correspond to the dimension-7 operators in \cref{eq:LagEFT1}. Assuming all dimensionless couplings are $\mathcal{O}(1)$, the energy scale of the effective operators is approximately given by
\begin{align}
    \Lambda_L \sim \left(\frac{m_{\phi}^2 m_{\phi_2}^2 }{m_{12}}\right)^{1/3} \,.
\end{align}
It was shown in \cref{fig:LambdaL} that to generate the observed active neutrino masses $\Lambda_L$ cannot be significantly larger than $\Lambda_5$. In the UV completion this corresponds to the requirement that $m_\phi \sim m_{\phi_2} \sim m_{12} \gtrsim \Lambda_5$.

\begin{table}[t]
  \renewcommand{\arraystretch}{1.5}
  \begin{ruledtabular}
    \begin{tabular}{lccccc}
      & $SU(5)$ & $SU(3)_c$ & $SU(2)_L$ & $U(1)_Y$ & $U(1)_\text{PQ}$ \\
      \colrule
      $\psi_{\bar{5}}$ & \Yvcentermath1$\overline{\tiny\yng(1)}$ & $\textbf{R}_\psi$ & $\mathbf{1}$ & $0$ & $-3/5$ \\
      $\psi_{10}$ & \Yvcentermath1${\tiny\yng(1,1)}$ & $\textbf{R}_\psi$ & $\mathbf{1}$ & $0$ & $1/5$ \\
      $\phi$ &\Yvcentermath1$\overline{\tiny\yng(1)}$ & $\textbf{R}_\psi$ & $\mathbf{1}$ & $0$ & $2/5$ \\
      $\phi_2$ & \Yvcentermath1$\overline{\tiny\yng(1)}$ & $\textbf{R}_\psi$ & \Yvcentermath1$\tiny\yng(1)$ & $-1/2$ & $2/5$ \\
      $L$ & $\mathbf{1}$  & $\mathbf{1}$ & \Yvcentermath1$\tiny\yng(1)$ & $-1/2$ & $1$ \\
      $H$ & $\mathbf{1}$  & $\mathbf{1}$ & \Yvcentermath1$\tiny\yng(1)$ & $1/2$ & $0$ \\
    \end{tabular}
  \end{ruledtabular}
  \caption{Representations of the fields in the UV completion of the heavy sterile neutrino scenario.}
  \label{tab:UVmodel-heavy}
\end{table}

\subsection{Light Sterile Neutrino Model}
\label{sec:lightsterile}

An alternative possibility is that the UV theory contains elementary, massless, right-handed neutrinos $\nu_R$ with PQ charge $-1$. The PQ symmetry forbids explicit Majorana mass terms for the $\nu_R$, but they form the usual Dirac masses with the $\nu_L$ (which here also have PQ charge $-1$). Tiny Yukawa couplings, $y_\nu$, would then normally be required to explain the active neutrino masses. However, the spontaneous PQ breaking in the $SU(5)$ sector can generate Majorana masses, $m_R$, for $\nu_R$, if there is mixing between the elementary $\nu_R$ and composite operators. As we shall show, $m_R$ can be hierarchically smaller than $\Lambda_5$, providing a means to naturally generate light sterile states and pseudo-Dirac neutrinos. 

\begin{figure*}[t]
    \centering
    \includegraphics[width = 0.49\linewidth]{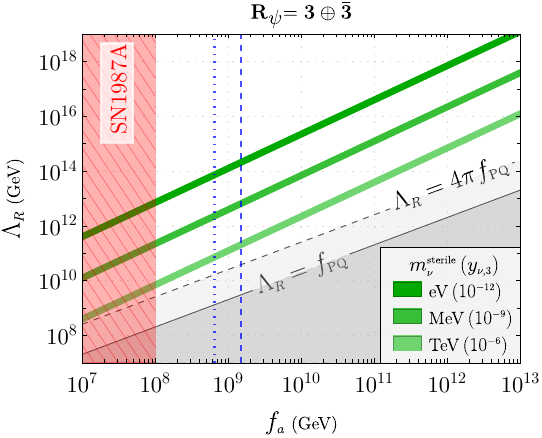} \hfill   
    \includegraphics[width = 0.49\linewidth]{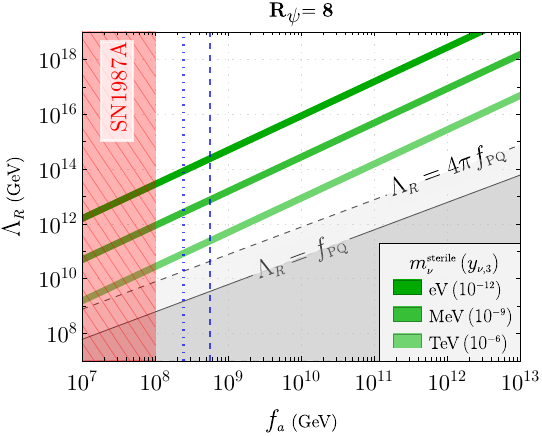}
    \caption{EFT scale $\Lambda_R$ versus $f_a$ in the light sterile neutrino model for $\mathbf{R}_\psi = \mathbf{3} \oplus \mathbf{\bar{3}}$ (left) and $\mathbf{R}_\psi =\mathbf{8}$ (right). The green bands depict contours of $m_\nu^{\text{sterile}}$, with the lower (upper) edges of the bands corresponding to $g_*=1\,(4\pi)$; the associated $y_{\nu,3}$ values for an active neutrino mass $m_{\nu,3}^{\text{active}} = 0.05\text{ eV}$ are shown in the legend. The red shaded region is excluded by the SN1987A bound on $f_a$~\cite{Chang:2018rso}. The estimated upper limit on $f_a$ for a high-quality axion consistent with the neutron EDM bound (assuming $\left| {\text{Im}}\left(c_{\cancel{\rm PQ}}\right)\right| \gtrsim 10^{-3}$, see \cref{app:axion-quality}) is shown by the blue dashed (dotted) line for $g_* = 1$ ($4\pi$). The breakdown of the EFT validity when $\Lambda_R \lesssim \Lambda_5 (=g_* f_{\rm PQ})$ is depicted by the dark (light) grey shaded region for $g_* = 1 \,(4\pi)$}
    \label{fig:LambdaR}
\end{figure*}

The elementary $\nu_R$ can mix with the 3-fermion, dimension-$\frac{9}{2}$ composite operators:
\begin{multline} \label{eq:LagEFT2}
    \mathcal{L}_\text{EFT} = \frac{\tilde{\zeta}_{ij}}{\Lambda_{R}^2} \nu_{R,i} \big( \psi_{\bar{5}}^\dagger  \psi_{\bar{5}}^\dagger \psi_{10}^\dagger \big)_j \\
    + \frac{\tilde{\zeta}'_{ij}}{\Lambda_{R}^2} \nu_{R,i} \big( \psi_{\bar{5}}^\dagger \psi_{10} \psi_{10} \big)_j 
    +\text{h.c.} \,,
\end{multline}
where $\Lambda_R$ is the EFT scale satisfying $\Lambda_5<\Lambda_R<M_\text{Pl}$, and $\tilde{\zeta}_{ij},\tilde{\zeta}'_{ij}$ are dimensionless couplings, with $i$ the neutrino flavor index and $j \in \{1,2\}$ enumerating the $SU(3)_c$ singlets in $(\psi\psi\psi)$. After the $SU(5)$ theory confines, these operators give rise to a (PQ-invariant) mass mixing between the $\nu_{R,i}$ and the composite fermions $N_j$. This effect is discussed further in \cref{app:simplemassmixing} in the context of a toy model with a single composite resonance. In the low-energy effective theory below the resonance scale $\Lambda_5$, the effect of the above operators is to generate a Majorana mass term for $\nu_R$. Including also the Dirac mass term for the neutrinos, we obtain
\begin{equation} \label{eq:numass2}
    \mathcal{L} = y_\nu^{ij} \frac{v}{\sqrt{2}} \,\nu^\dagger_{L,i} \nu_{R,j} + \frac{1}{2} m_R^{ij} \nu_{R,i} \nu_{R,j} e^{2ia/f_\text{PQ}}+ \text{h.c.} \,,
\end{equation}
with
\begin{equation}
    m_R^{ij} = 2\zeta_{ik} \zeta_{jk} \left(\frac{f_{\rm PQ}}{\Lambda_R}\right)^4 \Lambda_5 \,,
    \label{eq:mRmass}
\end{equation}
where $\zeta_{ij}\simeq\mathcal{O}(1)\times\tilde\zeta_{ij}$ and we have set $\tilde{\zeta}'_{ij}=0$ for simplicity. Notice that $m_R$ (which has again been estimated using dimensional analysis) is suppressed relative to the resonance scale, and if $\Lambda_R \gg f_{\rm PQ}$ there will be light sterile neutrino states. Hence, both the see-saw and pseudo-Dirac limits of the neutrino mass matrix in \eqref{eq:numass2} can be naturally obtained, depending on the ratio $f_\text{PQ}/\Lambda_R$. In the following, we assume diagonal couplings for simplicity: $y_\nu^{ij} = y_{\nu,i} \delta^{ij}$ and $\zeta_{ik}\zeta_{jk} = \delta_{ij}$. The see-saw limit of the neutrino masses is then given by
\begin{align}
    m_{\nu,i}^{\text{active}}&=y_{\nu,i}^2\left(\frac{\Lambda_R}{f_{\rm PQ}}\right)^4 \frac{v^2}{4\Lambda_5} \,, \label{eq:mnulightsterile}
\end{align}
and $m_\nu^{\text{sterile}}$ is given by \eqref{eq:mRmass}.

\Cref{fig:LambdaR} shows (green) contours of the sterile neutrino mass\footnote{Note that with our simplifying assumption $\zeta_{ik}\zeta_{jk} = \delta_{ij}$, the sterile masses are approximately degenerate.} in the $f_a$--$\Lambda_R$ plane. We have fixed $m_{\nu,3}^{\text{active}} = 0.05\,\text{eV}$, such that the contours correspond to different values of $y_{\nu,3}$. The left and right panels are for $\mathbf{R}_\psi = \mathbf{3} \oplus \mathbf{\bar{3}}$ and $\mathbf{R}_\psi =\mathbf{8}$, respectively. The red region is excluded by SN1987A and the region to the left of the blue lines is favoured to obtain a high-quality axion. The requirement that the EFT scale is above the resonance scale, $\Lambda_R \gtrsim \Lambda_5$, imposes an upper bound on $m_\nu^{\text{sterile}}$ (or equivalently $y_{\nu,3}$), as shown by the grey region.

The sterile neutrino masses can naturally be hierarchically smaller than the underlying scales $f_a$ and $\Lambda_R$. However, similar to the standard type-I seesaw, small Yukawa couplings, $y_\nu$, are then needed to obtain the active neutrino masses. With sterile neutrino masses of order the eV scale a coupling $y_\nu\sim 10^{-12}$ is needed, while TeV scale sterile neutrino states correspond to $y_\nu \sim 10^{-6}$. In the pure Dirac mass limit ($m_R\rightarrow 0$), an active neutrino mass of $m_\nu^\text{active}=0.05$ eV corresponds to $y_\nu \simeq 10^{-13}$.

\subsubsection{UV completion}

\begin{figure}[t]
    \centering
    \begin{subfigure}[t]{0.48\linewidth}
    \includegraphics[width = \linewidth]{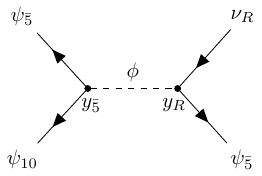}
    \caption{$\psi_{\bar 5}^\dagger \psi_{\bar 5}^\dagger\psi_{10}^\dagger\nu_R$}
    \end{subfigure}
    \begin{subfigure}[t]{0.48\linewidth}
    \includegraphics[width = \linewidth]{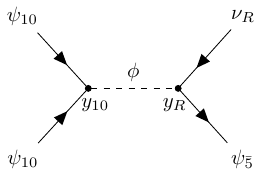}
    \caption{$\psi_{\bar 5}^\dagger \psi_{10}\psi_{10}\nu_R$}
    \end{subfigure}
    \caption{UV-completion of the dimension-6 operators in \cref{eq:UVlight} arising from tree-level exchange of the heavy scalar $\phi$. (The arrows represent fermion number.)} \label{fig:feynModelB}
\end{figure}

A tree-level UV completion for the operators in \cref{eq:LagEFT2} is obtained by introducing a massive, complex scalar field $\phi$ with mass $m_\phi > \Lambda_5$. Again, the scalar does not obtain a VEV and therefore does not affect the axion quality. The charge assignments of the fields are listed in \cref{tab:UVmodel-light}, leading to the interaction terms
\begin{multline}
    \mathcal{L} = y_\nu L^\dagger  H^\dagger \nu_R + y_{\bar{5}} \psi_{\bar{5}} \psi_{10} \phi + \frac{1}{2} y_{10} \psi_{10} \psi_{10} \phi^\dagger \\
    + y_R \nu_R \psi^\dagger_{\bar{5}} \phi + {\rm h.c.} \,,
    \label{eq:elemnuRLag}
\end{multline}
where $y_{\bar{5}},y_{10},y_R$ are dimensionless couplings and we have suppressed the indices that enumerate the $SU(3)_c$ contractions. For simplicity, we consider just one active neutrino flavor and one $\nu_R$ flavor. The analysis can be straightforwardly generalized to three active neutrino flavors. Integrating out the massive scalar $\phi$ in \eqref{eq:elemnuRLag}, as shown diagrammatically in \cref{fig:feynModelB}, yields
\begin{multline}
    \mathcal{L}_\text{eff} = \frac{y_{\bar{5}} y_R}{m_\phi^2} (\nu_R \psi^\dagger_{\bar{5}}) (\psi^\dagger_{\bar{5}} \psi_{10}^\dagger) \\
    + \frac{1}{2} \frac{y_{10} y_R}{m_\phi^2} (\nu_R \psi^\dagger_{\bar{5}}) (\psi_{10} \psi_{10}) + {\rm h.c.} \,, \label{eq:UVlight}
\end{multline}
which are the operators in \cref{eq:LagEFT2}, with $\Lambda_R \sim m_\phi$ for $\mathcal{O}(1)$ couplings.

\begin{table}[t]
  \renewcommand{\arraystretch}{1.5}
  \begin{ruledtabular}
    \begin{tabular}{lccccc}
      & $SU(5)$ & $SU(3)_c$ & $SU(2)_L$ & $U(1)_Y$ & $U(1)_\text{PQ}$ \\
      \colrule
      $\psi_{\bar{5}}$ & \Yvcentermath1$\overline{\tiny\yng(1)}$ & $\textbf{R}_\psi$ & $\mathbf{1}$ & $0$ & $-3/5$ \\
      $\psi_{10}$ & \Yvcentermath1$\tiny\yng(1,1)$ & $\textbf{R}_\psi$ & $\mathbf{1}$ & $0$ & $1/5$ \\
      $\phi$ & \Yvcentermath1$\overline{\tiny\yng(1)}$ & $\textbf{R}_\psi$ & $\mathbf{1}$ & $0$ & $2/5$ \\
      $\nu_R$ & $\mathbf{1}$ & $\mathbf{1}$ & $\mathbf{1}$ & $0$ & $-1$ \\
      $L$ & $\mathbf{1}$ & $\mathbf{1}$ & \Yvcentermath1$\tiny\yng(1)$ & $-1/2$ & $-1$ \\
       $H$ & $\mathbf{1}$ & $\mathbf{1}$ & \Yvcentermath1$\tiny\yng(1)$ & $1/2$ & $0$ \\
    \end{tabular}
  \end{ruledtabular}
  \caption{Representations of the fields in the UV completion of the light sterile neutrino scenario. (Note that $L$ has opposite PQ charge compared to the heavy sterile neutrino scenario.)}
  \label{tab:UVmodel-light}
\end{table}

\subsubsection{Holographic connection}

Our light sterile neutrino scenario provides a possible holographic realization of the 5D model considered in Ref.~\cite{Cox:2021lii}. In the 5D model, both the axion and right-handed neutrinos are bulk fields charged under a bulk $U(1)_{\rm PQ}$ gauge symmetry, which allows for hierarchically small sterile neutrino masses. According to the AdS/CFT dictionary (see e.g. Ref.~\cite{Gherghetta:2010cj}), the bulk $U(1)_{\rm PQ}$ gauge symmetry is dual to a global symmetry in the 4D gauge theory (CFT), while the Kaluza-Klein mass eigenstates can be understood as due to a mixing between an elementary and composite sector. 

For the right-handed neutrino, this would imply a mixing term $\nu_R {\cal O}$, where $\nu_R$ is an elementary fermion and $\cal O$ is an operator in the dual (CFT) gauge theory. A naturally small mixing can be generated when ${\rm dim}\,{\cal O}>4$. This is similar to what occurs in the light sterile neutrino case. This can be seen from the UV Lagrangian in \eqref{eq:UVlight}, where the operator is ${\cal O}=\psi^\dagger_{\bar{5}} \psi^\dagger_{\bar{5}} \psi^\dagger_{10}$ (assuming $y_{10}=0)$; the dual 4D theory is then identified as the $SU(5)$ gauge theory. Since ${\rm dim}\,{\cal O}=\frac{9}{2}$, the mixing is small (as seen in \cref{sec:lightsterile}) and therefore the sterile neutrino partner of the active neutrino can be naturally light. Thus, the UV completion considered in the light sterile neutrino case provides a specific holographic realization of the 5D model. 

This holographic realization is not in perfect agreement with the 5D model of Ref.~\cite{Cox:2021lii} because the left-handed (active) neutrinos were also bulk fields in the 5D model. This means that the dual theory should also feature mixing between the elementary $\nu_L$ and composite operators. It would be interesting to generalize our UV completion to also incorporate this feature. Finally, the axion in the 5D model has exponentially suppressed couplings on the UV brane. This corresponds to essentially a purely composite (and high-quality) axion in the dual theory, as also occurs in the $SU(5)$ gauge theory.

\section{Conclusion}
\label{sec:conc}

The axion and right-handed neutrinos are motivated by two seemingly unrelated puzzles of the Standard Model. In this work, we have provided a common origin for the QCD axion and right-handed neutrinos as bound states arising from strong dynamics. This builds upon the chiral $SU(5)$ gauge theory in Ref.~\cite{Gavela:2018paw}, which contains a high-quality composite axion, to also include composite neutrino states. This solution also provides a possible UV description of the holographic models considered in Refs.~\cite{Cox:2019rro,Cox:2021lii}.

Interestingly, the strong dynamics gives rise to composite sterile neutrino masses of order the PQ breaking scale. Depending on the origin of the coupling between the composite sterile neutrinos and the left-handed neutrinos, either pseudo-Dirac or Majorana neutrinos are possible. When the composite sterile neutrinos directly couple to left-handed neutrinos in a dimension-seven interaction with the Higgs, Majorana active neutrinos are obtained via a seesaw mechanism. The dimension-seven interaction can be generated by integrating out two PQ-charged, massive complex scalar fields in a UV completion that preserves the quality of the PQ symmetry. Alternatively, the composite sterile neutrinos can mix with elementary right-handed neutrinos, via dimension-$\frac{9}{2}$ operators, to induce naturally small couplings to the active neutrinos. This leads to sterile states that are hierarchically lighter than the PQ scale and can realise pseudo-Dirac neutrinos. The PQ symmetry plays the role of a generalised lepton number in realizing either the heavy or light sterile neutrino scenario. 

There are number of phenomenological features of our model that could be tested in future experiments. In the pseudo-Dirac limit there is a contribution to the number of effective neutrino species; for reheating temperatures above the $SU(5)$ confining phase transition, $\Delta N_{\rm eff}\sim 0.1$, which can be tested in upcoming CMB experiments~\cite{Abazajian:2019oqj,Adshead:2020ekg,Chakraborty:2021fkp}. Alternatively, if the light sterile neutrinos have sub-TeV masses they could be detected at collider experiments. In the post-inflationary scenario (assuming the residual discrete PQ symmetry is broken to avoid stable domain walls), a first order $SU(5)$ phase transition could give rise to a gravitational wave signal associated with the PQ scale (see e.g.~\cite{VonHarling:2019rgb}), which is worthy of further study. Our model also predicts axion-neutrino couplings that could lead to effects in neutrino oscillations within the local DM axion halo~\cite{Gherghetta:2023myo}. Finally, baryogenesis mechanisms can be straightforwardly incorporated into our model, such as the usual leptogenesis mechanism, or a cogenesis mechanism, as considered in Ref.~\cite{Chakraborty:2021fkp}. 

The coincidence between the axion decay constant and seesaw mass scales can therefore be explained by strong dynamics, which naturally connects the axion and neutrinos in a way that can also address the axion quality problem.


\begin{acknowledgments}
P.C. is supported by the Australian Research Council Discovery Early Career Researcher Award DE210100446. The work of T.G. was supported in part by DOE grant DE-SC0011842 at the University of Minnesota. A part of this work was performed by T.G. at the Aspen Center for Physics, which is supported by National Science Foundation grant PHY-1607611.
\end{acknowledgments}


\appendix

\section{QCD Anomaly Factor}
\label{app:QCDanom}

The anomalous non-conservation of the PQ current due to the $U(1)_\text{PQ} \times SU(3)_c^2$ anomaly is
\begin{align}
    \partial_\mu j^\mu_{\text{PQ}} = \mathcal{N}\,\frac{\alpha_s}{8\pi} G\widetilde{G} 
    \label{eq:PQ-QCD-anomaly}\,,
\end{align}
where $\widetilde G_{\mu\nu}=\frac{1}{2}\varepsilon_{\mu\nu\rho\sigma}G^{\rho\sigma}$ and $\varepsilon_{0123}=+1$. As usual, the QCD anomaly factor $\mathcal{N}$ can be obtained from the one-loop triangle diagram that connects the PQ current with two gluon fields:
\begin{align}
    &\mathlarger{\mathlarger{\sum}}_\psi~ \begin{gathered}\includegraphics[scale=.7]{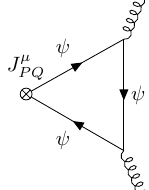} \end{gathered} ~~\sim~ \sum_\psi Q_\psi^{\text{PQ}}\cdot 2T(\mathbf{R}_\psi)\,,
    \label{eq:33barN}
\end{align}
where $Q_\psi^{\text{PQ}}$ is the PQ charge and $T(\mathbf{R}_\psi)$ is the index of the $SU(3)_c$ representation.

We consider two possibilities for the (real) representation $\textbf{R}_\psi$ of the fermions:  $\mathbf{R}_\psi = \textbf{3} \oplus \overline{\textbf{3}}$ and the adjoint representation $\mathbf{R}_\psi = \textbf{8}$.  For the \texorpdfstring{$\mathbf{3} \oplus \overline{\mathbf{3}}$}{3 + 3bar} model we obtain
\begin{align}
    \mathcal{N} = &\sum_\psi Q_\psi^{\text{PQ}}\cdot 2T(\mathbf{R}_\psi)\,,\nonumber\\
    &= 5 Q_{\bar{5}}\cdot 2 \left( T(\mathbf{3}) + T(\overline{\mathbf{3}}) \right) + 10 Q_{10}\cdot 2 \left( T(\mathbf{3}) + T(\overline{\mathbf{3}}) \right)\,, \nonumber \\
    &= -2\,, 
\end{align}
where we have used the PQ charges in Table~\ref{tab:global} and $T(\mathbf{3})=T(\overline{\mathbf{3}})=\frac{1}{2}$.

Similarly, for the color octet model we have
\begin{align}
    \mathcal{N} =  &\sum_\psi Q_\psi^{\text{PQ}}\cdot 2T(\mathbf{R}_\psi)\,, \nonumber\\
    &=  5 Q_{\bar{5}}\cdot 2T(\mathbf{8})  + 10 Q_{10}\cdot 2T(\mathbf{8})= -6\,,
    \label{eq:8N}
\end{align}
using $T(\mathbf{8})=3$.

\section{\texorpdfstring{$SU(3)_c$ Representations of NG Bosons}{SU(3)c Representations of NG Bosons}}
\label{app:NGB-rep}

In this Appendix, we discuss the $SU(3)_c$ representations of the NG bosons of the $SU(n_f)_{\bar{5}}\times SU(n_f)_{10}$ flavor symmetry breaking. The NG bosons are parameterised in terms of the broken generators of the flavor group; hence, we require the decomposition of the adjoint representation of the flavor group under the $SU(3)_c$ subgroup. This depends on the embedding of $SU(3)_c$ in the flavor group. 

It can be seen from \cref{tab:gauge,tab:global} that the fundamental representations of the $SU(n_f)_{\bar{5}}$ and $SU(n_f)_{10}$ flavor groups decompose into representations of the $SU(3)_c$ subgroup as 
\begin{align}
    \tiny\yng(1)\, \rightarrow \,\mathbf{R}_\psi\,.
    \label{eq:fund-decomp}
    \end{align}
The same decomposition holds for the anti-fundamental representations, since $\mathbf{R}_\psi$ is a (pseudo-) real representation of $SU(3)_c$. 

The adjoint representations of the $SU(n_f)$ flavor groups can be constructed from the tensor product of the fundamental and the anti-fundamental representations:
\begin{align}
    \tiny\yng(1)\,\otimes\,\overline{\yng(1)} = \mathbf{Adj}\oplus \mathbf{1} \,.
    \label{eq:adjrep}
\end{align}
Then, using \eqref{eq:fund-decomp} we see that the decomposition of the adjoint into $SU(3)_c$ representations is contained within the $SU(3)_c$ tensor product $\mathbf{R}_\psi \otimes \mathbf{R}_\psi$. For the two specific representations we consider, $\mathbf{R}_\psi = \mathbf{3}\oplus\overline{\mathbf{3}}$ and $\mathbf{R}_\psi = \mathbf{8}$, this is
\begin{align}
    (\mathbf{3}\oplus\overline{\mathbf{3}}) \otimes (\mathbf{3}\oplus\overline{\mathbf{3}}) &= 2(\mathbf{8})\oplus \mathbf{6}\oplus \overline{\mathbf{6}}\oplus\mathbf{3}\oplus \overline{\mathbf{3}}\oplus 2(\mathbf{1}) \,,\label{eq:tensorDecom3} \\
    \mathbf{8} \otimes \mathbf{8} &= \mathbf{27}\oplus \mathbf{10}\oplus \overline{\mathbf{10}}\oplus 2(\mathbf{8})\oplus \mathbf{1} \,. \label{eq:tensorDecom8}
\end{align}
Comparing (\ref{eq:tensorDecom3}) and (\ref{eq:tensorDecom8}) with (\ref{eq:adjrep}), we can obtain the decomposition of the adjoint representation of the full $SU(n_f)_{\bar{5}}\times SU(n_f)_{10}$ flavor group under the $SU(3)_c$ subgroup for the two cases:
\begin{itemize}
    \item $\mathbf{R}_\psi = \mathbf{3}\oplus\overline{\mathbf{3}}$:
    \begin{equation}
        \mathbf{Adj}\,\rightarrow\, 2(2(\mathbf{8})\oplus \mathbf{6}\oplus \overline{\mathbf{6}}\oplus\mathbf{3}\oplus \overline{\mathbf{3}}\oplus \mathbf{1}) \,. \label{eq:embedding3}
    \end{equation}
    \item $\mathbf{R}_\psi = \mathbf{8}$:
    \begin{equation}
        \mathbf{Adj}\,\rightarrow\,2(\mathbf{27}\oplus \mathbf{10}\oplus \overline{\mathbf{10}}\oplus 2(\mathbf{8})) \,. \label{eq:embedding8}
    \end{equation}
\end{itemize}

One of the adjoint representations (i.e.\,$\mathbf{8}$) of $SU(3)_c$ in the RHS of both (\ref{eq:embedding3}) and (\ref{eq:embedding8}) corresponds to the generators of the unbroken $SU(3)_c$ subgroup. The remainder gives the $SU(3)_c$ representations of the NG bosons of the spontaneous symmetry breaking $SU(n_f)_{\bar{5}}\times SU(n_f)_{10} \to SU(3)_c$ (i.e. when the group $\mathcal{G}$ in \cref{eq:global-symmetry-breaking} is trivial). Importantly, as discussed in \cref{sec:symmetry-breaking}, all of the NG bosons are colored in the $\mathbf{R}_\psi = \mathbf{8}$ case and hence obtain masses of order $\sqrt{\alpha_s} \Lambda_5$. For $\mathbf{R}_\psi = \mathbf{3}\oplus\overline{\mathbf{3}}$, there are two massless, QCD singlet NG bosons. This corresponds to the fact that, in this case, the gauging of $SU(3)_c$ preserves two residual $U(1)$ flavor symmetries. Due to the cosmological bounds on additional relativistic degrees of freedom, this scenario is only viable if the composite sector remains out of equilibrium with, and colder than, the SM bath. On the other hand, if these $U(1)$ symmetries are instead contained in $\mathcal{G}$ (i.e. not spontaneously broken by the strong dynamics) then there are no QCD singlet NG bosons, which would alleviate the cosmological requirements.

\section{Explicit PQ Breaking and Axion Quality}
\label{app:axion-quality}

A feature of the $SU(5)$ chiral gauge theory is that the PQ symmetry is accidentally preserved up to very high order~\cite{Gavela:2018paw}. As discussed in \cref{sec:symmetry-breaking}, the lowest dimension $SU(5)$ and $SU(3)_c$ gauge invariant, Lorentz scalar operators that have non-zero PQ charge contain six fermion fields. This implies that the PQ symmetry is \textit{accidentally} preserved up to (gravitationally-induced) dimension nine terms in the Lagrangian. The relevant operators are listed in \cref{eq:PQcondensate}. 

While the leading PQ-breaking operators contain six-fermion fields in the present $SU(5)$ model, it is interesting to consider whether there are generalisations of this model in which PQ-breaking terms arise from eight-fermion (or higher) operators, since this would provide a more robust solution to the axion quality problem. A simple extension is to consider $SU(N)$ gauge groups with $N \geq 6$ and fermions in the antifundamental and antisymmetric irreps. Using the Mathematica package \textit{GroupMath}~\cite{Fonseca:2020vke}, we find that there always exist PQ-breaking operators with either four or six fermions for all $6\leq N\leq 16$.

\subsection{Displacement of the axion potential minimum}

Planck suppressed, PQ-charged operators cause a displacement of the axion potential minimum from its CP conserving value. To determine this displacement, we consider the following Lagrangian containing the (leading) dimension nine  operators,
\begin{align}
    \mathcal{L}_{\cancel{\rm PQ}} = \frac{1}{4\pi}\frac{1}{M_{\rm Pl}^5}\,\Bigg[&\frac{c_1}{2!\,4!}\, \psi_{\bar{5}}\,\psi_{\bar{5}}\,\psi_{10}\,\psi_{\bar{5}}\,\psi_{\bar{5}}\,\psi_{10} \notag \\
    + &\frac{c_2}{2!\,4!}\, \psi_{\bar{5}}\,\psi_{10}^\dagger\,\psi_{10}^\dagger\,\psi_{\bar{5}}\,\psi_{10}^\dagger\,\psi_{10}^\dagger \notag \\
    + &\frac{c_3}{2!\,3!}\, \psi_{\bar{5}}\,\psi_{\bar{5}}\, \psi_{10}\,\psi_{\bar{5}}\,\psi_{10}^\dagger\,\psi_{10}^\dagger \notag \\
    + &\frac{c_4}{4!}\, \psi_{\bar{5}} \,\psi_{\bar{5}}\,\psi_{\bar{5}}\,\psi_{\bar{5}}\,\psi_{\bar{5}}^\dagger\,\psi_{10}^\dagger\Bigg]+\text{h.c.}\,,
    \label{eq:PQbreakingL}
\end{align}
where the $c_{i}$ are dimensionless constants and the overall prefactor has been estimated using naive dimensional analysis (NDA)~\cite{Manohar:1983md,Gavela:2016bzc}, assuming the gravitational EFT scale $\Lambda_{\rm Pl}=4\pi M_{\rm Pl}$ and $M_{\rm Pl}\approx 10^{19}\,$GeV. All of the above operators have PQ charge $-2$ and each term represents multiple gauge singlet combinations. For example, the decomposition of each operator into irreducible representations of $SU(5)$ is
\begin{align}
    \Phi_{\rm PQ,1} 
    \rightarrow &\,\bm{\bar{5}} \otimes \bm{\bar{5}}  \otimes \bm{10} \otimes \bm{\bar{5}} \otimes \bm{\bar{5}} \otimes \bm{10} = 6(\bm{1}) \oplus ... \,, \label{eq:dim6-1}
    \\
    {\Phi}_{\rm PQ,2} 
    \rightarrow & \,\bm{\bar{5}} \otimes \bm{\overline{10}} \otimes \bm{\overline{10}} \otimes \bm{\bar{5}} \otimes \bm{\overline{10}} \otimes \bm{\overline{10}} = 9(\bm{1}) \oplus ... \,, \label{eq:dim6-2}
    \\
    {\Phi}_{\rm PQ,3} 
    \rightarrow &\,\bm{\bar{5}} \otimes \bm{\bar{5}}  \otimes \bm{10} \otimes \bm{\bar{5}} \otimes \bm{\overline{10}} \otimes \bm{\overline{10}} = 7(\bm{1}) \oplus ... \,,  \label{eq:dim6-3}
       \\
    {\Phi}_{\rm PQ,4} 
    \rightarrow & \,\bm{\bar{5}} \otimes \bm{\bar{5}}  \otimes \bm{\bar{5}} \otimes \bm{\bar{5}} \otimes \bm{5} \otimes \bm{\overline{10}} = 4(\bm{1}) \oplus ... \,, \label{eq:dim6-4}
\end{align}
showing that $\Phi_{\rm PQ,1}$ includes six $SU(5)$ invariant combinations, and similarly for $\Phi_{\rm PQ,2}$, $\Phi_{\rm PQ,3}$ and $\Phi_{\rm PQ,4}$. There is an analogous decomposition for $SU(3)_c$, further increasing the number of gauge singlet combinations.

The Lagrangian \eqref{eq:PQbreakingL} gives rise to the following term in the low-energy effective theory below the $SU(5)$ resonance scale,
\begin{equation}
    \mathcal{L}_{\cancel{\rm PQ}} = \frac{c_{\cancel{\rm PQ}}}{4\pi}\,\frac{(\Lambda_5 f_{\rm PQ}^2)^3}{M_{\rm Pl}^5}
    e^{-2i\frac{a}{f_{\rm PQ}}}+\text{h.c.} \,,\label{eqn:PQbreakingL2}
\end{equation}
where $c_{\cancel{\rm PQ}}$ is a constant that depends on the $c_{i}$ in \eqref{eq:PQbreakingL} and is assumed to be $\mathcal{O}(1)$. The resulting axion potential is then approximately given by
\begin{align} \label{eq:axion-potential}
    V(a) \approx -m_a^2 & f_a^2 \cos\left(\frac{a}{f_a}\right) \nonumber \\ 
    &-2|c_{\cancel{\rm PQ}}|\frac{g_*^3}{4\pi}\frac{\mathcal{N}^9f_a^9}{M_{\rm Pl}^5}\, \cos\left(\frac{2}{\mathcal{N}}\frac{a}{f_a} - \delta\right) \,,
\end{align}
where $\mathcal{N}$ is the QCD anomaly factor (see \cref{app:QCDanom}), and $f_a\equiv f_{\rm PQ}/\mathcal{N}$. The first term is the usual QCD contribution and the second term arises from \eqref{eqn:PQbreakingL2}, with $c_{\cancel{\rm PQ}}=|c_{\cancel{\rm PQ}}|\,e^{i\delta}$ and $\delta$ representing an arbitrary phase from gravity that is not necessarily aligned with the phases in the SM.

\begin{figure}[t]
 \centering
    \includegraphics[width = \linewidth]{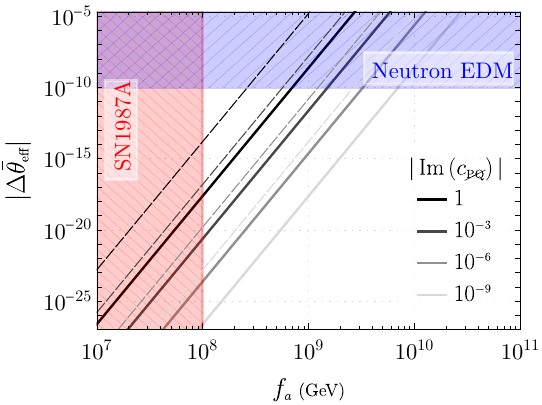}
    \caption{Displacement from the CP conserving minimum, $|\Delta \bar{\theta}_{\text{eff}}|$, due to Planck suppressed operators as a function of $f_a$. The contours show different values of $\left| {\text{Im}} \left(c_{\cancel{\rm PQ}}\right)\right|$, assuming $g_*=1$. The solid (dashed) lines correspond to the QCD representation $\mathbf{R}_\psi = \mathbf{3} \oplus \mathbf{\bar{3}}$ ($\mathbf{R}_\psi =\mathbf{8}$). The blue and red regions are excluded by the upper bound on the neutron EDM and SN1987A, respectively.\label{fig:fa-theta}}
\end{figure}

The displacement of the axion potential minimum with respect to the CP conserving minimum is then found to be
\begin{equation}
    |\Delta \bar{\theta}_{\text{eff}}| \approx \left| {\text{Im}}\left(c_{\cancel{\rm PQ}}\right)\right| \frac{g_*^3}{4\pi} \frac{\mathcal{N}^8f_a^7}{M_{\rm Pl}^5\,m_a^2}\,.
    \label{Deltheta}
\end{equation}
This displacement is constrained by the upper bound on the neutron EDM to be $|\Delta \bar{\theta}_{\text{eff}}| \lesssim 10^{-10}$ ~\cite{Abel:2020pzs}. \Cref{fig:fa-theta} shows the displacement $|\Delta \bar{\theta}_{\text{eff}}|$ as a function of $f_a$ for several values of $\left| {\text{Im}}\left(c_{\cancel{\rm PQ}}\right)\right|$. Taking into account the lower bound on $f_a$ from SN1987A~\cite{Chang:2018rso}, we see that a solution to the axion quality problem with $\left| {\text{Im}}\left(c_{\cancel{\rm PQ}}\right)\right| \in (0.001,\,1)$ requires $10^8 \lesssim f_a/\text{GeV} \lesssim 6\times  10^8\,(2\times 10^8)$ for QCD representations $\mathbf{R}_\psi =\mathbf{ 3\oplus \bar{3}}$ ($\mathbf{R}_\psi =\mathbf{8}$) and $g_*=1$. These bounds are similar to those given in Ref.~\cite{Gavela:2018paw}; however, we have included all the dimension nine PQ-breaking operators in our analysis.

\section{Axion Dark Matter}
\label{app:dark-matter}

As is well-known, the axion provides one of the best-motivated dark matter candidates. The production of axion dark matter in the early universe depends on whether the PQ-breaking occurs before or after the inflationary era.

If PQ-breaking occurs before (or during) inflation, then axion dark matter is produced via the misalignment mechanism. The relic axion abundance $\Omega_a$ is then given by~\cite{Ballesteros:2016xej,Borsanyi:2016ksw}
\begin{equation}
    \Omega_a h^2=0.12\,{\theta^2_i} \left(\frac{f_a}{9\times 10^{11}\text{GeV}}\right)^{1.165} \,, \label{misalign}
\end{equation}
where $\theta_{i}=a_i/f_a$ is the initial misalignment angle, and $h\simeq 0.68$ is the present-day Hubble parameter (in units of 100 km s$^{-1}$ Mpc$^{-1}$). If the total dark matter relic density, $\Omega_{\text{DM}}\,h^2\simeq 0.12$~\cite{Planck:2018vyg}, is due to axions, the required range of $f_a$ for an initial misalignment angle in the range $\theta_{i}\in (0.1,3)$ is
\begin{equation}
    10^{11}\,\text{GeV} \lesssim f_a \lesssim 5\times 10^{13} \,\text{GeV} \,. 
    \label{fadmpreferred}
\end{equation}
This range is in tension with realizing the PQ symmetry as a high-quality accidental symmetry of the $SU(5)$ chiral gauge theory.

A modification of the misalignment mechanism is to assume that the initial velocity of the axion is nonzero -- the so-called kinetic misalignment mechanism~\cite{Co:2019jts}. For an elementary axion, this mechanism can produce the correct relic abundance for any decay constant in the range $10^8~{\rm GeV}\lesssim f_a\lesssim 1.5 \times 10^{11}$~GeV. It would be interesting to explore whether this mechanism can be implemented in a composite axion scenario, particularly for values of the decay constant $f_a\sim 10^8$\,GeV that ameliorate the axion quality problem. Both of the above scenarios assume that the universe is {\it not} reheated to temperatures above the $SU(5)$ de-confinement transition, which would restore the PQ symmetry.

Alternatively, in the post-inflationary PQ-breaking scenario the universe is reheated to temperatures above the PQ-breaking scale. As the universe cools, topological defects form which, depending on the domain wall number, may include stable domain walls. To determine the domain wall number, we first note that the QCD contribution to the axion potential in \cref{eq:axion-potential} preserves the discrete symmetry,
\begin{align}
    \frac{a}{f_a} \rightarrow \frac{a}{f_a} + 2\pi n,~~n\in \mathbb{Z}\,.
\end{align}
The physical domain of the axion field is $a/f_a \in [0,{2\pi|\mathcal{N}|})$, with the anomaly coefficient $\mathcal{N}=-2$ and $\mathcal{N}=-6$ for the $\mathbf{R}_\psi = \mathbf{3} \oplus \overline{\mathbf{3}}$ and $\mathbf{R}_\psi = \mathbf{8}$ models, respectively (see \cref{app:QCDanom}). Therefore, the number of degenerate minima of the QCD potential in the physical domain, which is equivalent to the number of domain walls, is
\begin{align}
    N_{\text{DW}} = 
    \begin{cases}
        2 & \quad \mathbf{R}_\psi = \mathbf{3} \oplus \overline{\mathbf{3}}\,,\\
        6 & \quad \mathbf{R}_\psi = \mathbf{8}\,.
    \end{cases}
    \label{NDW}
\end{align}

Since $N_{\rm DW}>1$, explicit violation of the discrete PQ symmetry, which would allow domain walls to decay, is required if the post-inflationary scenario is to be viable. In principle, such a ``bias'' potential~\cite{Sikivie:1982qv} which lifts the vacuum degeneracy could simply arise from Planck-suppressed, higher-dimension operators that explicitly violate the PQ symmetry, as considered in \eqref{eq:dim6-1}-\eqref{eq:dim6-4} or of higher dimension. For instance, 6 fermion, dimension-9 terms reduce the domain wall number in the octet model to $N_{\rm DW}=2$. However, going beyond 6-fermion terms does not reduce the domain wall number further and therefore a new source of breaking would be required to lift the remaining degeneracy. This may arise, for example, if $SU(3)_c$ is embedded in a larger gauge group, such as recently considered in Ref.~\cite{Cordova:2023her}.

Assuming such a bias potential, the decays of cosmic strings and domain walls contribute to the axion dark matter density. With the present state-of-the-art calculations (see e.g. Refs.~\cite{Gorghetto:2020qws,Buschmann:2021sdq} for the case of axion strings and Ref.~\cite{Gorghetto:2022ikz} for $N_{\rm DW}>1$), there remains significant uncertainty in the quantitative estimation of the axion abundance from the decay of topological defects, which can dominate over the misalignment contribution. Therefore, the lower bound on $f_a$ arising from the dark matter relic density in the pre-inflationary scenario (in \cref{fadmpreferred}) could be significantly relaxed in the post-inflationary scenario. In fact, a robust upper bound of $f_a\lesssim 5.4\times 10^8$ GeV~ (or  $m_a\gtrsim 11$ meV)~\cite{Beyer:2022ywc}, can be derived on obtaining the required axion relic abundance from domain wall decay.

\section{Mass Mixing in the Light-Sterile Neutrino Model}
\label{app:simplemassmixing}

In this Appendix, we present a toy model of the mass mixing in the light-sterile neutrino scenario by including a single right-handed neutrino resonance $N_1$ in the effective low-energy theory. The Lagrangian for this simplified model of the effects of the strong dynamics is given by
\begin{align} \label{eq:neutL}
     \mathcal{L} &\simeq y_\nu \frac{v}{\sqrt{2}} \,\nu^\dagger_L \nu_R  + \frac{1}{2} \Lambda_5 N_1 N_1 \nonumber \\
     &\qquad\qquad\qquad + \frac{1}{2} y_{\bar 5}y_R \frac{\Lambda_5 f_{\rm PQ}^2}{m_\phi^2} {\nu}_R N_1 + \text{h.c.} \\
     & \equiv \renewcommand{\arraystretch}{1}
     \frac{1}{2}\begin{pmatrix}\nu^\dagger_L & \nu_R & N_1 \end{pmatrix}\begin{pmatrix}0 & m_D & 0\\m_D & 0 & \Delta_R\\
     0 & \Delta_R &m_N\\\end{pmatrix} \begin{pmatrix}{\nu^\dagger_L} \\ \nu_R
     \\N_1\end{pmatrix} + \text{h.c.} \,,
     \label{eq:massmatrx}
\end{align}
where $\Delta_R$ is a Dirac mass mixing between the elementary field $\nu_R$ and the composite resonance $N_1$, whose value depends on the parameters of the UV completion. Note that in terms of the constituent fields, the operator corresponding to the Majorana mass term for $N_1$ is ${\Phi}_{\rm PQ,1}$ in \cref{eq:PQcondensate}, which can be split into two PQ-charged fermionic operators.

\begin{figure}[H]
    \centering
    \includegraphics[width = \linewidth]{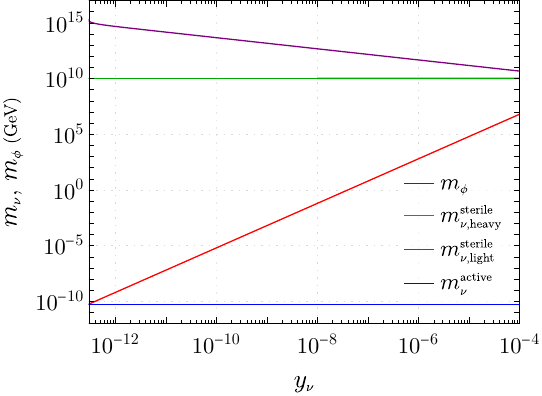}
    \caption{Neutrino and scalar masses as a function of $y_\nu$, assuming an active neutrino mass $m_\nu^\text{active}=0.05\,\text{eV}$, $g_*= 1$, $y_{\bar{5}} = y_R = 1$, and $f_{\rm PQ}=10^{10}\,\text{GeV}$. The Yukawa coupling is restricted to $y_\nu \lesssim 10^{-4}$ in order to satisfy $m_\phi \gtrsim \Lambda_5$.} \label{fig:numassvsy}
\end{figure}

The neutrino mass eigenvalues of \eqref{eq:massmatrx} are determined numerically and shown in Fig.~\ref{fig:numassvsy} as a function of the Yukawa coupling $y_\nu$, assuming $f_{\rm PQ}=10^{10}\,\text{GeV}$, $g_*=1$, and with the active neutrino mass set to $0.05\,\text{eV}$. In the pure Dirac mass limit for the active neutrinos ($\Delta_R\rightarrow 0$), a mass of $m_\nu^\text{active}=0.05$ eV corresponds to $y_\nu \simeq 10^{-13}$. As the mass $m_\phi$ of the scalar field in the UV completion (or $\Lambda_R$ in the EFT) is lowered, the sterile partner of the active neutrino increases in mass via the mixing with $N_1$. More generally, the sterile partner of the active neutrino will mix with all the resonances of the strong dynamics.
\vspace{1em}

\bibliography{SU5_axion}

\end{document}